\documentclass[]{JHEP3}   % 10pt is ignored!

\JHEPspecialurl{http://jhep.sissa.it/JOURNAL/JHEP3.tar.gz}

\usepackage{epsfig,multicol}

\newcommand\fverb{\setbox\pippobox=\hbox\bgroup\verb}
\newcommand\fverbdo{\egroup\medskip\noindent%
            \fbox{\unhbox\pippobox}\ }
\newcommand\fverbit{\egroup\item[\fbox{\unhbox\pippobox}]}
\newbox\pippobox

\title{
\begin{flushright}
\normalsize{ FERMILAB-Pub-06/066-T\\FTUV-06/0329}
\end{flushright}
Minimal Noncanonical Cosmologies}

\author{       {Gabriela Barenboim}\\
\normalsize\emph{Departament de F\'isica Te\`orica, Universitat de
Val\`encia}\\ 
\emph{Carrer Dr. Moliner 50, E-46100 Burjassot (Val\`encia), Spain}\\
Email: \email{gabriela.barenboim@uv.es}\\}
\author{\textbf{Joseph D. Lykken}\\
\normalsize\emph{Fermi National Accelerator Laboratory}\\
\emph{P.O. Box 500, Batavia, IL 60510, USA}\\
Email: \email{lykken@fnal.gov}}

\abstract{We demonstrate how much it is possible
to deviate from the standard cosmological paradigm of inflation-assisted
$\Lambda$CDM, keeping within
current observational constraints, and {\it without adding to or
modifying any theoretical assumptions.}
We show that within a minimal framework there are
many new possibilities, some of them wildly different from the standard
picture. We present three illustrative examples of new models,
described phenomenologically by a noncanonical scalar field coupled to
radiation and matter. These models have interesting implications
for inflation, quintessence, reheating, electroweak baryogenesis, 
and the relic densities of WIMPs and other exotics.}


\begin{document}

\newcommand{\be}{\begin{eqnarray}}
\newcommand{\ee}{\end{eqnarray}}
%%% ----------------------------------------------------------------------

\section{Introduction}

The recent release of the WMAP three year data \cite{Spergel:2006hy}
illustrates the extent to which cosmological model-building is now
contrained and guided by precision data. It also emphasizes the extent
to which a Standard Model of cosmology has emerged. This concordance
model is denoted the ``power--law $\Lambda$CDM model'' in \cite{Spergel:2006hy};
a slightly more theory--loaded designation is the ``inflation--assisted
$\Lambda$CDM model''. In this model the universe on large scales is
homogeneous, isotropic, and spatially flat. The universe contains
matter (dominated by dark matter), radiation, and dark energy. The dark energy
consists either of a cosmological constant or of a quintessence field
whose equation of state parameter $w(t)$ is close to $-1$ now. 
Primordial density
perturbations have a nearly scale invariant spectrum. At scales
comparable to the current Hubble radius, the scalar perturbations have
a slightly negative spectral tilt \cite{Spergel:2006hy}.

A weakness of the cosmological data set is that direct observations
only cover the era in which the scale factor $a(t)$ was about $0.001$ or
larger. The success of Big Bang Nucleosynthesis (BBN) suggests that
the the universe was strongly radiation-dominated at $a \simeq 10^{-10}$,
and that not much entropy was added to the radiation bath after the
time of BBN. Beyond this, and especially in order to describe earlier 
times, we have to add additional theoretical baggage.
A primordial epoch
of inflaton-driven accelerated expansion provides an attractive
(though not unique) explanation for many features of cosmological
data. Thus the assumption of an inflaton seems like a minimal
theoretical input allowing us to make models of early time
cosmology that can be confronted with data.  

Having allowed for the possibility of an inflaton and a quintessence field,
it is important to ask: what is the general class of such models currently
allowed by data? The purpose of this paper is to begin to answer this
question.We will do this in a purely phenomenological framework. Thus we
avoid the impossible task of trying to classify all possible 
additional top--down theoretical assumptions, and are able to discuss
a wide variety of models in a single framework.

Our aim is to demonstrate how much it is possible
to deviate from the standard cosmological paradigm, keeping within
current observational constraints, and {\it without adding to or
modifying any theoretical assumptions.} 
We will show that within a simple framework there are
many new possibilities, some of them wildly different from the standard
picture. Since these possibilities do not involve any new theoretical
assumptions or inputs, Occam's razor does not separate them from
the standard paradigm.

To be completely concrete, let us restate the theoretical assumptions 
that underlie standard cosmology:
\begin{enumerate}
\item The large scale evolution of the universe is well-described by 
solutions of the four dimensional Einstein equations which are spatially
homogeneous and isotropic (FRW).
This assumption holds for any prior epoch
when the temperature is (very roughly) less than
$M_{Planck}$. During this time the universe has
been expanding monotonically.
\item The primordial distribution of light nuclei was produced by
the standard BBN processes.
\item The solution to the horizon and flatness problems is
provided by inflationary expansion in earlier epochs. This
process is parametrized by a single scalar inflaton, which also
produces primordial scalar and curvature perturbations which
are nearly scale invariant, consistent
{\it e.g.}, with observations of the Cosmic Microwave Background
(CMB).  
\item The dominant form of matter now is cold dark matter, 
consistent with a total $\Omega$ indistinguishable from unity,
and thus with the above assumption of primordial inflation.
\item Large scale evolution has lately undergone a transformation
from a matter-dominated expansion to an accelerating expansion.
The source of this accelerating expansion is parametrized either by
a small positive cosmological constant, or by a quintessence scalar
whose equation of state parameter $w(t)$ is close to $-1$ now.
\end{enumerate}

We regard the above assumptions as minimal, and will refer to
them in this paper as {\it the minimal standard assumptions}.
However we wish to attack
the usual conclusion, {\it i.e.} that the
minimal standard assumptions imply a cosmological history
of the universe essentially identical to the standard picture of
inflation--assisted $\Lambda$CDM cosmology. Our purpose is to demonstrate
that current data allow a much richer range of possibilities, even
within this conservative set of assumptions. We will exhibit several
examples, all within an identical framework. One of these examples
is superficially similar to the standard picture, but differs
in one important feature.
The other two are qualitatively
very different from the standard picture and from each other.
Future observations, or better analysis of existing data,
may be able to distinguish between these alternative histories.

In the next section we describe a general framework for constructing
models which satisfy the minimal standard assumptions. This framework
parametrizes FRW cosmological histories in terms of a single scalar
field $\theta$ with a potential $V(\theta )$ and (in general) a
noncanonical kinetic function $F(\theta )$. In section 3 we compare this
framework to a minimal approach to inflation, in which an inflaton
scalar drives a primordial epoch of accelerated expansion, and 
generates a nearly scale invariant spectrum of scalar and
curvature perturbations. As widely acknowledged, without further
theoretical assumptions cosmological data do not yet distinguish
whether the inflaton is a fundamental degree of freedom or merely
an effective description of other physical processes. Thus
we propose a much larger class of cosmological histories, that
can likewise be described phenomenologically as driven by
a scalar field. This larger class is constrained by data but involves
no new kinds of theoretical inputs. Any particular model is fully specified
by $V(\theta )$ and $F(\theta )$ combined with some prescription for
how the $\theta$ field couples to ordinary radiation and matter.

In section 4 we describe three examples of new models which obey 
the minimal standard assumptions, described phenomenologically in
the noncanonical framework. The first model resembles the
standard picture, with both the inflaton and the quintessence scalar
identified with $\theta$. There is a long epoch of primordial inflation,
enough to solve the horizon and flatness problems. However
inflation is less rapid than
in the standard picture, and the temperature 
decreases much more slowly,
obviating the need for a period of strong reheating. The second example
is the ``slinky'' model already presented in \cite{Barenboim:2005np}.
In this model a somewhat abbreviated epoch of primordial inflation
is followed by a second period of inflation which begins just before
the electroweak phase transition (EWPT), and ends before BBN. 
Again the radiation temperature
decreases much more slowly than in the standard picture, and there is
no period of large reheating. The Hubble expansion rate is much smaller
during the EWPT; for Higgs sectors such that the 
phase transition is first order,
this will enhance electroweak baryogenesis.
In the third example there are many inflationary epochs;
the current accelerated expansion marks the beginning of the sixth
period of inflation. The fifth period of inflation occurs after
BBN, but without upsetting the baryon to photon ratio. Other periods
of inflation occur before and after the EWPT, 
also affecting the relic density of
WIMPs, gravitinos, and modulinos, if present.

\section{A noncanonical framework for FRW cosmology}

It is not widely appreciated that any FRW cosmological history can
be parametrized by a single real scalar field $\theta (x^{\mu})$
coupled to Einstein
gravity. In general the scalar field action will be noncanonical,
meaning that it has the form
\begin{eqnarray}
\label{eqn:gennocanaction}
\int d^4x \; \sqrt{-g}\; \left[  
\frac{1}{2}\,F(\theta )\,P(X)
-V(\theta ) \right] \; ,
\end{eqnarray}
where $X=\frac{1}{2}g^{\mu\nu}\partial_{\mu}\theta\partial_{\nu}\theta$.
When $V=0$ and $P(0)=0$, this is the class of noncanonical
theories which generate k--essence models 
of inflation \cite{Chiba:1999ka}-\cite{Malquarti:2003nn}.
When $P(X)=X$, as will be true in all of our examples, 
(\ref{eqn:gennocanaction}) reduces to
\begin{eqnarray}
\label{eqn:nocanaction}
\int d^4x \; \sqrt{-g}\; \left[  
\frac{1}{2}\,F(\theta )\,
g^{\mu\nu}\partial_{\mu}\theta\partial_{\nu}\theta 
-V(\theta ) \right] \; .
\end{eqnarray}
Of course one can transform such noncanonical scalars into a canonical ones
via a field redefinition, but this is not useful if (as occurs in ours
models) the vacuum expectation value of $F(\theta )$ vanishes at
certain times.

An FRW cosmological history is completely specified by a
scale factor $a(t)$, where without loss of generality we can
take the scale factor now to be unity: $a_0 = 1$. Thus the
cosmological history is equivalently specified by the Hubble
parameter $H(t) \equiv \dot{a}/a$. Since we assumed that the
FRW expansion is monotonic, we can trade the co-moving time
variable $t$ for the scale factor $a$. Thus an FRW cosmological
history is fully specified by some $H(a)$.

Given any such history, we can immediately write down a noncanonical
scalar theory which reproduces it:
\begin{eqnarray}
\label{eqn:thetarel}
\theta (a) &=& -b\,{\rm ln}\, a \; ,\\
\label{eqn:Frel}
F(\theta ) &=& \frac{3k^2}{2b^2}(1+w(\theta )) \; ,\\
\label{eqn:Vrel}
V(\theta ) &=& \frac{3k^2}{4}(1-w(\theta ))H^2 \; ,
\end{eqnarray}
where $k^2 = M_{Planck}^2/4\pi$, and 
$b$ is an arbitrary real parameter introduced for convenience.
The equation of state parameter $w(\theta )$ of the scalar is given by:
\begin{eqnarray}
\label{eqn:wrel}
1+w(\theta ) = \frac{2b}{3}\frac{H'}{H} \; ,
\end{eqnarray}
where the prime denotes a derivative with respect to $\theta$,
using the relation (\ref{eqn:thetarel}).

It is easy to show that (\ref{eqn:thetarel})-(\ref{eqn:wrel})
define a simultaneous solution to the Friedmann equation, the
equation of motion of the scalar, and the continuity equation:
\begin{eqnarray}
H^2 &=& \frac{2}{3k^2}\rho_{\theta} \; ,\\
0 &=& \ddot\theta + 3H\dot\theta +\frac{1}{F}
(\frac{1}{2}{\dot\theta}^2F' + V') \; , \\
{\dot\rho}_{\theta} &=& -3(1+w)H\rho_{\theta} \; ,
\end{eqnarray}
where $\rho_{\theta}$ is the scalar energy density. Note that, for convenience,
we have defined $\theta$ in (\ref{eqn:thetarel}) such that
$\theta = 0 $ now.

This framework is best understood by working out some
familiar examples:

\subsection{$\mathbf{\Lambda}$CDM cosmology}
It is trivial to reproduce pure $\Lambda$CDM cosmology
using this framework. Our input is an FRW cosmological history
defined by:
\begin{eqnarray}
H^2(a) = \frac{2}{3k^2}\left(
\rho_{\Lambda}^0 +\rho_m^0\,a^{-3} + \rho_r^0\, a^{-4} \right) \; ,
\end{eqnarray}
where $\rho_{\Lambda}^0$, $\rho_m^0$ and $\rho_r^0$ are constants.
This cosmology is reproduced by a noncanonical scalar field
theory with $F(\theta )$ and $V(\theta )$ given by:
\begin{eqnarray}
F(\theta ) &=& \frac{3k^2}{2b^2}\;
\frac{\rho_m^0{\rm e}^{3\theta /b} + \frac{4}{3}\rho_r^0
{\rm e}^{4\theta /b}}{\rho_{\Lambda}^0 + \rho_m^0{\rm e}^{3\theta /b}
+\rho_r^0{\rm e}^{4\theta /b}} \; ,\\
V(\theta ) &=& \rho_{\Lambda}^0 + \frac{1}{2}\rho_m^0{\rm e}^{3\theta /b}
+\frac{1}{3}\rho_r^0{\rm e}^{4\theta /b} \; .
\end{eqnarray}
These functions have a rather peculiar form, but this is not
surprising since we are mocking up an FRW expansion which in reality
is driven by many different components.

\subsection{ultra-slow roll inflation}

Standard inflationary models address an earlier FRW epoch,
in which a long inflationary phase with $H(a) \sim $ constant
somehow exits gracefully into a reheated radiation dominated phase.
If the scalar field interactions responsible for reheating are
perturbative, they can be modeled within the FRW formalism 
by perturbative ``friction'' in the coupled equations of motion
of the inflation, radiation and matter. If the scalar field
interactions are nonperturbative, as in the parametric resonances
responsible for preheating \cite{Kofman:1994rk}-\cite{Dufaux:2006ee},
then our simple framework is not adequate,
and needs to be expanded to incorporate this additional dynamics.

During the inflationary phase, scalar and curvature perturbations
are produced, with power spectra that are close to scale-invariant.
Ultra--slow roll inflation  \cite{Tsamis:2003px}-\cite{Kinney:2005vj}
is a toy model for this behavior,
in which the inflaton potential is a constant, and the late-time
limit of the FRW evolution is a pure de Sitter phase.
Ultra--slow roll 
inflation is specified by a single free parameter $V_0$:
\begin{eqnarray}
H^2(a) = \frac{2V_0}{3k^2}\,(1+a^6) \; .
\end{eqnarray}
In our general framework this corresponds to:
\begin{eqnarray}
F(\theta ) &=& \frac{3k^2}{2b^2}{\rm e}^{3\theta /b}{\rm cosh}\,
(3\theta /b) \; ,\\
V(\theta ) &=& V_0 \; .
\end{eqnarray}    
The ``slow roll'' parameters $\epsilon$, $\eta$ and $\xi$ can be computed
in the noncanonical framework using the results of \cite{baren-kinney}:
\begin{eqnarray}
\label{eqn:nonceps}
\epsilon &=& k^2\frac{1}{F}\left(\frac{H'}{H}\right)^2 
= 3\frac{a^6}{1+a^6} \; ,\\
\label{eqn:nonceta}
\eta &=& \frac{k^2}{F}\left[
\left(\frac{H'}{H}\right)^2+\frac{1}{2}\frac{F'H'}{FH} \right]
= 3
\; ,\\
\xi^2 &=& 3(\epsilon + \eta ) -\eta^2
-\frac{1}{FH^2}\left( V'' -\frac{1}{2}\frac{F'}{F}V' \right)
= 3\epsilon \; ,
\end{eqnarray}
where in each line the second equality reproduces the standard results
for ultra--slow roll inflation.

\subsection{hybrid inflation}

Hybrid inflation \cite{Linde:1993cn}
is usually described by two canonical scalar fields
$\phi (x)$ and $\psi (x)$ with a potential given by
\be
V(\phi ,\psi ) = \left( M - \frac{\sqrt{\lambda}}{2}\psi^2 \right)
+\frac{1}{2}m^2\phi^2 + \frac{1}{2}g\phi^2\psi^2 \; .
\ee
Hybrid inflation has 
two phases \cite{Garcia-Bellido:1996ke},\cite{Kinney:1997ne}. 
In the first phase, inflation
occurs while $\psi = 0$ and the $\phi$ inflaton rolls towards
a critical value $\phi_c$, given by
\be
\phi_c^2 = \frac{2\sqrt{\lambda}}{g}M^2 \; .
\ee
During this inflationary phase the potential reduces to
\be\label{eqn:simplehypot}
V(\phi ) = M^4 + \frac{1}{2}m^2\phi^2 \; .
\ee
During the second phase of hybrid inflation, the $\psi$ field becomes
unstable and inflation ends. If $\lambda \gg g^2$, then reheating
proceeds perturbatively through oscillations 
of the $\phi$ field \cite{Garcia-Bellido:1997wm}.
Thus this case is amenable to description by our simple framework
with a single inflaton. Here we will only explicitly reproduce 
the first inflationary phase.

With these caveats, the FRW cosmological history of hybrid inflation
is determined once we specify the parameters $\epsilon$ and $\eta$:
\be\label{eqn:hypars}
\epsilon = \frac{r_{\pm}^2}{k^2}\,\phi^2 \quad ;\qquad
\eta = \epsilon + r_{\pm} \; ,
\ee
where there are two branches of solutions given by
\be
r_{\pm} = \frac{3}{2}\left[
1\mp\sqrt{1-\frac{2}{3}\left(\frac{m^2k^2}{M^4}\right)}
\right] \; .
\ee

In our framework this is recovered by the choices:
\be\label{eqn:ourhyF}
F(\theta ) &=& \frac{r_{\pm}^2}{b^2}\,{\rm e}^{2r_{\pm}\theta /b} \; ,\\
\label{eqn:ourhyV}
V(\theta ) &=& M^4\left( 1 - \frac{1}{3}\frac{r_{\pm}^2}{k^2}\,
{\rm e}^{2r_{\pm}\theta /b}
\right) {\rm exp}\left(\frac{r_{\pm}^2}{k^2}\,{\rm e}^{2r_{\pm}\theta /b}\right)
\; .
\ee
The potential has a rather strange form. However, in computing the
power spectra of density perturbations, we are always interested
in the case 
\be
\frac{r_{\pm}^2}{k^2}\,{\rm e}^{2r_{\pm}\theta /b} \ll 1 \; ,
\ee
in which case both the $r_+$ and $r_-$ branches reduce to
\be
V(\theta ) = M^4 + \frac{1}{2}m^2{\rm e}^{2r_{\pm}\theta /b} 
\; ,
\ee
as expected from (\ref{eqn:simplehypot}).

It is easy to verify using the first equalities of
(\ref{eqn:nonceps}), (\ref{eqn:nonceta}) that (\ref{eqn:ourhyF}),
(\ref{eqn:ourhyV}) produce the slow roll parameters of hybrid
inflation as given in (\ref{eqn:hypars}).
In hybrid inflation one is in the slow roll regime if \cite{Kinney:1997ne}:
\be
\frac{m^2k^2}{M^4} \ll 1 \; .
\ee
We have not made this assumption; more generally, our framework is not
tied to the slow roll approximation.

\section{What is the inflaton?}

The general framework just described ties into
a key dilemma of inflationary theory. While it is certainly possible
to generate a variety of phenomenologically interesting 
FRW cosmological histories from an inflaton scalar, this in itself
gives us no idea how seriously to take this scalar as a
fundamental degree of freedom. Examining the inflation potential
is of no help. If the potential has a simple form, then
we must certainly worry that it could be a stand--in for other
physical mechanisms. If the inflaton potential is complicated,
then we also suspect that it may be a stand--in for a combination
of effects, as was the case in our $\Lambda$CDM example of the
previous section.  

Reheating is a much more promising guide, since it depends on how the inflaton
couples to other degrees of freedom, and to radiation in particular.
Evidence for very efficient reheating or preheating could be
smoking guns for resonances or other specific dynamical features.
But even with a single inflaton
scalar, there is a lot of freedom to mock up a variety of mechanisms
which transfer energy back and forth between the scalar and
radiation or matter. So we are a still a long way from
being able to use this information (ultimately derived from data)
to reveal the true degrees of freedom behind the inflaton.

The inevitable conclusion is that presently we have little guidance for
how to distinguish which inflaton potentials are more
``plausible'' on theoretical grounds. Instead the current thrust of
inflationary theory is properly towards the bottom--up 
approach of reconstructing purely
phenomenological inflaton potentials 
from data \cite{Copeland:1993zn}-\cite{Cline:2006db}.
The framework that we are promoting here is a useful
middle ground for linking this bottom--up approach to
top--down models -- models that begin with more narrow assumptions
about new fundamental degrees of freedom.

From the discussion of the previous section, it is clear that
many cosmologies that satisfy the minimal standard assumptions
can be represented by a single noncanonical scalar coupled to
radiation and matter. This includes standard
inflation models that match onto $\Lambda$CDM, as well as some
models of quintessential inflation.
A model is specified by the functions $F(\theta )$, $V(\theta )$,
and by some phenomenological friction terms in the equations
of motion, which represent the interactions of this scalar with
radiation and matter. Current data, and the minimal standard
assumptions themselves, place rather strong constraints on which
models of this type are viable.

In this general framework, as we have already seen, there is
little motivation to regard some models as simpler or more plausible
than others. All models that satisfy the constraints from data
are equally good, in the sense that they have the
same number of degrees of freedom and satisfy the same
set of theoretical assumptions. We can only distinguish among
models by adding more theoretical assumptions, or adding
further constraints from data.

Models of this general class are generically models of quintesential
inflation; this is to say that $w$ of the noncanonical scalar is
close to $-1$ today, and $\dot w$ is nonzero over most or all of 
FRW cosmological history. This kind of scenario was the motivation
for the ``slinky'' model of quintessential inflation proposed
in \cite{Barenboim:2005np}, as well as 
the ``undulant'' cosmologies discussed 
in \cite{Barenboim:2004kz},\cite{Barenboim:2005wx}.
In fact the entire class of models
that we are describing can be considered as a generalization of
the particular slinky model presented in \cite{Barenboim:2005np}.
This model gives a very nonstandard cosmological
history. This is our first clue that there is a rich class of new models
that follow from the minimal standard assumptions.

In the next section we present three examples of new models. All of them
obey the minimal standard assumptions. The first gives an FRW cosmological
history which looks fairly standard, but with a simple picture for
quintessence and a gentle approach to reheating. 
The second and third examples are very different from
the standard picture, but appear to satisfy all the obvious constraints
from data.

\section{New models}

The models are based on the following forms for
$F(\theta )$ and $V(\theta )$:
\begin{eqnarray}\label{eqn:ourkin}
F(\theta ) &=& \frac{12k^2}{b^2}\, {\rm sin}^2\,\theta\; ;\\
V(\theta ) &=& \rho_0\, {\rm cos}^2\,\theta\, {\rm exp}\left[{3\over b}\left(
2\theta - {\rm sin}\,2\theta \right) \right] \; ,
\label{eqn:ourpot} 
\end{eqnarray}
where $\rho_0$ is fixed to be the dark energy density today.
These choices are motivated by asking for a simple periodic
behavior in the the equation of state parameter:
\begin{eqnarray}
w(a) = -{\rm cos}\,2\theta (a) \; .
\end{eqnarray}
Of course we could attempt a simpler form for $V(\theta )$ at the
expense of getting a more complicated form for $w(a)$, but this gets
us into the kind of top-down model-building that we are trying (for
the moment) to avoid. 

Our simple framework assumes that the coupling between
quintessence and radiation or matter can be described
perturbatively. The simplest modification of the
evolution equations consistent with energy
conservation is: 
\begin{eqnarray}
\dot{\rho}_{\theta} &=& -3H(1+w)\rho_{\theta} 
-k_0m_{\phi}(1+w)\rho_{\theta}
\; ,\nonumber\\
\dot{\rho}_r &=& -4H\rho_r 
+(1\hspace*{-2pt}-\hspace*{-2pt}f_m)k_0m_{\phi}(1+w)\rho_{\theta}
\; ,\\
\dot{\rho}_m &=& -3H\rho_m +f_mk_0m_{\phi}(1+w)\rho_{\theta}
\; ,\nonumber
\label{eqn:coupled}
\end{eqnarray}
where $k_0$ and $f_m$ are small dimensionless functions of time.
In our examples these will just be constants, or constants
multiplied by step functions.
As long as $\theta$ is not near
a multiple of $\pi/2$, it is a reasonable approximation to make
the replacement \cite{Barenboim:2005np}
\begin{eqnarray}
k_0m_{\phi} \to kH \; ,
\label{eqn:makeghappy}
\end{eqnarray}
where $k$ is another small dimensionless parameter.
This replacement decouples the $\theta$ evolution equation
from the Friedmann equation, giving an immediate analytic
solution:
\begin{eqnarray}
\rho_{\theta}(a) = \rho_0 \, 
{\rm exp}\left[{1\over b}(3+k)\left(
2\theta - {\rm sin}\,2\theta \right) \right]
\; .
\label{eqn:rhonew}
\end{eqnarray}
We have used this convenient but nonessential
approximation in generating the figures shown below.

Models of this type thus have four adjustable parameters:
$b$, $k$, $f_m$, and $a_i = a(t_i)$, with $t_i$ the initial time at
which we begin the FRW evolution. Two of these parameters,
$k$ and $f_m$, are essentially fixed by requiring that the
matter and radiation fractions today agree with data.
We do not distinguish between the excess of baryonic matter density
and the much larger excess of dark matter density, thus
$f_m$ refers to the production of dark matter. Here we are
assuming that the dark matter results from decay of 
the noncanonical scalar. If instead the dark matter is a
thermal relic, then one can set $f_m =0$ and obtain the
dark matter relic density by standard methods. Note however
that such calculations must take into account the nonstandard
expansion histories in our models.

\begin{figure}
\centerline{\epsfxsize 5.8 truein \epsfbox {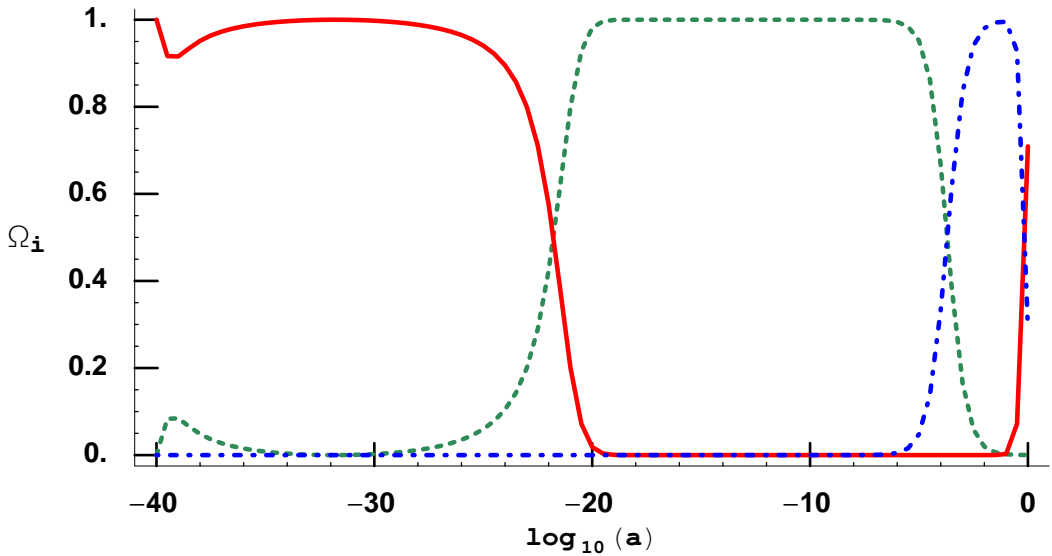}}
\caption{\label{fig:modelone} The cosmological history of
Model 1. Shown are the relative energy density fractions
in radiation (green/dashed), matter (blue/dot--dashed), and
the noncanonical scalar (red/solid), as a function of the
logarithm of the scale factor.
\hbox to170pt{}}
\end{figure}

The remaining parameters $b$ and $a_i$ are constrained
by several phenomenological requirements:
\begin{itemize}
\item Since we assume standard BBN to explain the abundances
of light nuclei, the universe should be radiation dominated when
the temperature is in the MeV range. In addition, we should not
produce very much entropy, in the form of radiation from
reheating, at any time after BBN.
\item To solve the horizon problem, the ratio of the comoving
horizon to the comoving Hubble radius, as measured today,
should be greater than one:
\be
aH\int_0^a \frac{da'}{a'}\,\frac{1}{a'H(a')} > 1 \; .
\ee
\item The scalar spectral index, for perturbations
which are now on scales comparable to the Hubble radius,
should be close to the WMAP--preferred value \cite{Spergel:2006hy}.
\item The temperature $T = (30\rho_r/\pi^2g_*)^{1/4}$ should
not exceed $M_{Planck}$ at any time after $t_i$.
\item Late time variations of $w(a)$ should not interfere
with structure formation, or cause too much distortion of
the imprint from baryon acoustic oscillations (BAO).
\item The coupling of the scalar to radiation and ordinary
matter must be very suppressed at late times, to satisfy bounds
on Equivalence Principle violations, Faraday rotation of
light from distant sources, and time variation of Standard
Model parameters \cite{Carroll:1998zi},\cite{Feldman:2006wg}. 
\end{itemize}

While $b$ and $a_i$ are nontrivially constrained by the
above considerations, many solutions remain. The tunings
required are not very strong; all three of our examples were
obtained from a few minutes of hand-tuning, not from a
systematic parameter scan.

\subsection{A simple model of quintessential inflation with gentle
reheating}

This model is defined by:
\be
b=1/11.7 \; ,\quad k=0.14\;\theta (10^{-10}-a)\; ,
\quad f_m = 3\times 10^{-20}\; ,
\quad a_i = 10^{-40} \; ,
\ee
where the step function $\theta (10^{-10}-a)$ turns off the coupling of the
scalar to radiation and matter from BBN time until now. 
This choice is motivated by
the need to suppress the coupling of the scalar to Standard Model 
fields at late times,
but this particular implementation is just an example from a class 
of similar models. 
The parameter $f_m$ is very small, as is to be expected. It was 
already noted above that
we can set $f_m = 0$ if we have in mind thermal relic dark matter 
of some specified
variety.

This model resembles the standard paradigm in many respects. The 
FRW history begins
with a single long period of inflation. This crosses over to a long epoch of
radiation domination beginning at a temperature of about $10^9$ GeV. 
At late times,
before recombination, a matter dominated phase begins. A new accelerated
expansion is beginning now. The horizon problem is solved because   
the ratio of the comoving
horizon to the comoving Hubble radius, as measured today,
is about 3.

\begin{figure}
\centerline{\epsfxsize 5.0 truein \epsfbox {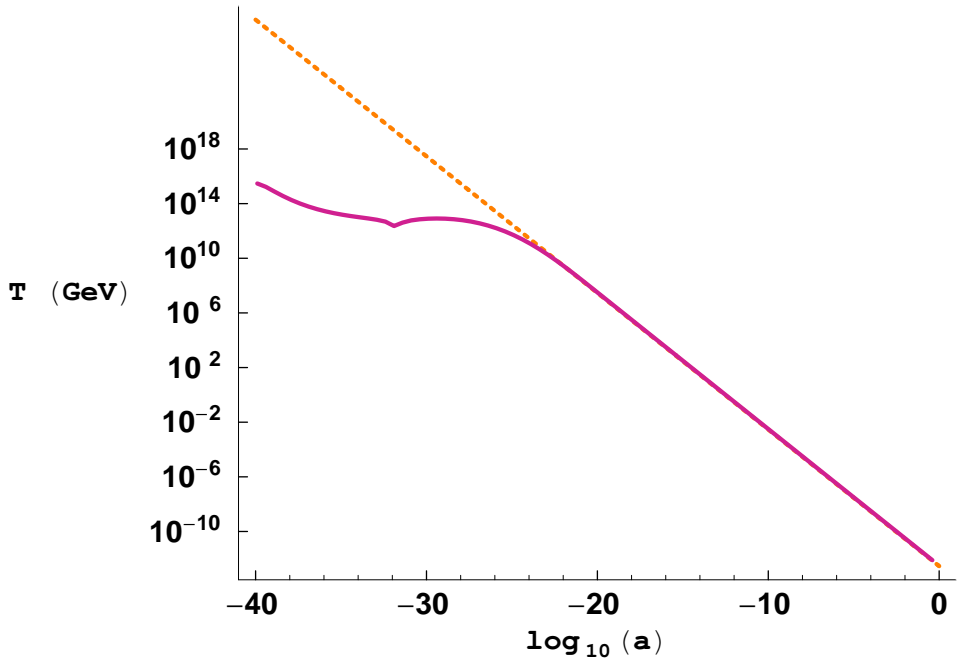}}
\caption{\label{fig:modelonetemp} The temperature history of
Model 1 (violet/solid), compared to a purely radiation-dominated cosmology
(yellow/dashed).
\hbox to170pt{}}
\end{figure}

On the other hand, the temperature history of this model is quite
nonstandard. Due to the relatively large value of the parameter $k$,
radiation and the noncanonical scalar track \cite{Zlatev:1998tr}
each other.
The maximum temperature is about $2\times 10^{15}$ GeV.
As can be seen from Figure \ref{fig:modelonetemp}, there is no dramatic
reheating phase. Instead, the temperature falls off very slowly during
the long inflationary phase, due to the tracking behavior.

It is difficult to extract a precise prediction for
the spectral indices of this model (or the two following),
since we are never strictly in the slow roll regime.
We will be content here with a rough estimate.
This is obtained starting from a canonical field
redefinition:
\be
\phi (x) = 2\sqrt{3}\frac{k}{b}\,{\rm cos}\,\theta \; .
\ee
In terms of the canonical scalar $\phi$, the potential
\ref{eqn:ourpot} can be written:
\be
V(\phi ) &\propto&  \phi^2 \;  H^2 (\phi ) \; ,
\ee
where $H(\phi )$ is the Hubble rate we would get
ignoring radiation and matter.
During inflation, $H$ is approximately constant,
but this is not an especially good approximation since
we are not in a slow roll regime. This is similar to
the oscillatory models of quintessential inflation
discussed in \cite{Nunes:2002wz}. Taking the potential
in the inflationary phase to be approximated by
$V \propto \phi^2$, we can estimate the
scalar spectral index $n_s$:
\begin{eqnarray}
n_s  \simeq 1 -\frac{2}{N}
\end{eqnarray}
with $N$ the number of e-folds between the Hubble radius exit and
the end of the inflationary period.
For our model $ N = 29$, giving $n_s = .93$, in good agreement with
recent observations \cite{Spergel:2006hy}.

For the two models following, inflation takes place in
installments rather than during a single primordial period.
However the total number of e-folds
of inflation remains approximately the same, so we expect
the spectral indices to be roughly the same.

\begin{table}[tb]
\centering
\begin{tabular}{|c|c|c|c|c|c|c|}
\hline\hline
log$_{10}$ $a $ & -40 & -38 & -36 & -34 & -32 & -30 \\ [0.5 ex]
\hline
$w(a)$ & 0.02 & -0.37 & -0.70 & -0.92 & -1.0 & -0.93 \\ \hline
$T$ & 0 & 2 10$^{14}$ & 3 10$^{13}$ & 10$^{13}$ & 3 10$^{12}$ & 8 10$^{12}$ \\ \hline
$\Omega_\Lambda$ & 1 & 0.95 & 0.98 & 0.997 &
0.99997 & 0.998 \\ \hline
$\Omega_r$ & 0 & 0.05 & 0.02 & 0.003 & 0.00003 & 0.002 \\ \hline
$\Omega_m$ & 0 & 4 10$^{-21}$ & 7 10$^{-22}$ & 10$^{-22}$
& 2 10$^{-24}$ & 10$^{-22}$   \\\hline
log$_{10}$ $a $ & -28 & -26 & -24 & -22 & -20 & -18 \\\hline
$w(a)$ & -0.72 & -0.39 & -0.01 & 0.37 & 0.70 & 0.92 \\ \hline
$T$ & 6 10$^{12}$ & 2 10$^{12}$ &  10$^{11}$ & 3 10$^{9}$ & 3 10$^{7}$ & 3 10$^{5}$  \\ \hline
$\Omega_\Lambda$ & 0.99 & 0.96 & 0.9 & 
0.6 & 0.02 & 7 10$^{-6}$ \\\hline
$\Omega_r$ & 0.01 & 0.04 & 0.1 & 0.4 & 0.98 & 0.999993 \\\hline
$\Omega_m$ & 5 10$^{-22}$ & 2 10$^{-21}$ & 10$^{-20}$
& 3 10$^{-19}$ & 5 10$^{-17}$ & 5 10$^{-15}$   \\\hline
log$_{10}$ $a $ & -16 & -14 & -12 & -10 & -8 & -6 \\\hline
$w(a)$ & 1.0 & 0.93 & 0.71 & 0.39 & 0.004 & -0.38 \\ \hline 
$T$ & 3000 & 30 & 0.3 & 3 10$^{-3}$ & 3 10$^{-5}$ & 3 10$^{-7}$  \\ \hline
$\Omega_\Lambda$ & 3 10$^{-10}$ & 10$^{-14}$ & 4 10$^{-18}$ & 6 10$^{-20}$ & 2 10$^{-19}$
& 10$^{-16}$ \\\hline
$\Omega_r$ & 1.0 & 1.0 & 1.0 & 1.0 & 0.99995 & 0.995  \\\hline
$\Omega_m$ & 5 10$^{-13}$ & 5 10$^{-11}$ & 5 10$^{-9}$
& 5 10$^{-7}$ & 5 10$^{-5}$ & 0.005 \\\hline
log$_{10}$ $a $ & -4 & -2 & 0 &  &  &  \\\hline
$w(a)$ & -0.71 & -0.92 & -1.0 &  &  &  \\ \hline
$T$ & 3 10$^{-9}$ & 3 10$^{-11}$ & 3 10$^{-13}$ &  &  &   \\ \hline
$\Omega_\Lambda$ & 10$^{-11}$ & 3 10$^{-6}$ & 0.71 &  &  &   \\\hline
$\Omega_r$ & 0.67 & 0.02 & 0.00006 &  &  &  \\\hline
$\Omega_m$ & 0.33 & 0.98 & 0.29 & &  &  \\[1ex]
\hline\hline
\end{tabular}
\caption{Model 1: the relative density fractions of dark energy, radiation,
and matter, as a function of the scale factor. Also shown are the
temperature $T$ in GeV, and the equation of state parameter $w(a)$.}
\end{table}

\subsection{A model with an extra inflationary epoch}

This model is defined by:
\be
b=1/7 \; ,\quad k=0.058\; ,\quad f_m = 10^{-11}\; ,
\quad a_i = 10^{-42} \; .
\ee
This is the slinky model of \cite{Barenboim:2005np}, with a slight
decrease in the parameter $k$ to get a better fit to the WMAP
preferred value for $\Omega_{\Lambda}$.

\begin{figure}
\centerline{\epsfxsize 5.5 truein \epsfbox {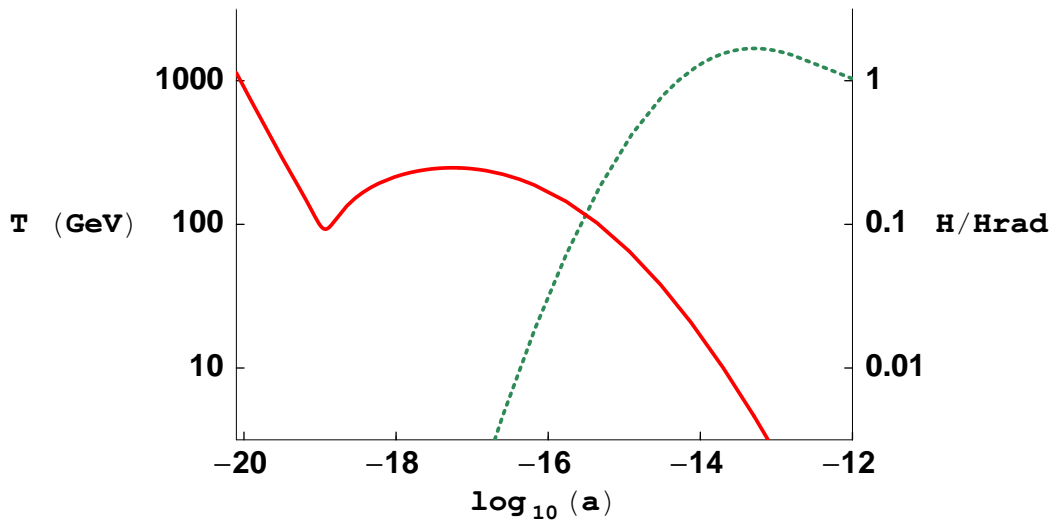}}
\caption{\label{fig:modeltwotemp} The temperature history of
Model 2 (red/solid) near the electroweak phase transition. Shown in
green/dashed is the Hubble parameter $H$ of Model 2 normalized to
the expansion rate $H_{rad}$ for pure radiation.
\hbox to170pt{}}
\end{figure}

In this model, we are currently beginning the third epoch of accelerated
expansion. A second period of accelerated expansion began just before
the electroweak phase transition, and ended well before BBN.
The temperature history near the EWPT is shown in Figure \ref{fig:modeltwotemp}.
Also shown is the Hubble parameter $H$ of Model 2 normalized to
the expansion rate $H_{rad}$ for pure radiation. $H_{rad}$ corresponds
to what is assumed in the standard paradigm. Notice that for temperatures
of a few GeV the expansion rate is actually somewhat larger than normal,
but at higher temperatures it is much less than normal.

Such a nonstandard thermal history will impact on electroweak baryogenesis.
For a Higgs sector such that the EWPT is first order, the change in the net
baryon asymmetry is proportional to -log$(H/H_{rad})$, where $H$
is the expansion rate during the phase transition, and $H_{rad}$ is
the corresponding expansion rate for pure radiation \cite{Joyce:1997fc}.
If the Higgs sector is such that the EWPT is second order, the
baryon asymmetry is proportional to the expansion rate \cite{Joyce:1997fc}.
Clearly one should revaluate the popular scenarios
for electroweak baryogenesis \cite{Trodden:1998ym} in this light.

\begin{figure}
\centerline{\epsfxsize 5.8 truein \epsfbox {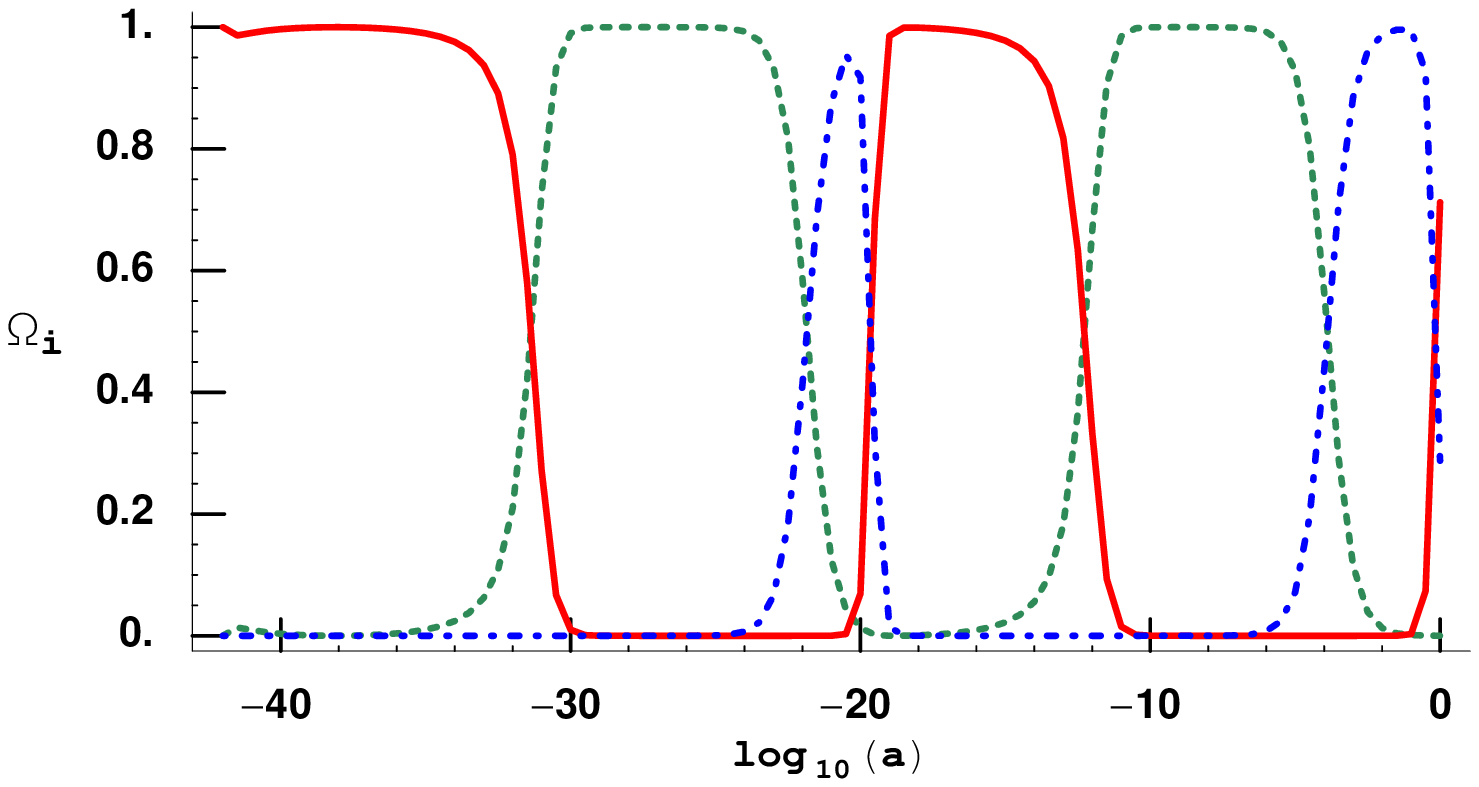}}
\caption{\label{fig:modeltwo} The cosmological history of
Model 2. Shown are the relative energy density fractions
in radiation (green/dashed), matter (blue/dot--dashed), and
the noncanonical scalar (red/solid), as a function of the
logarithm of the scale factor.
\hbox to170pt{}}
\end{figure}

Model 2 will have major implications for 
predictions of the relic abundance of dark matter particles with Terascale
masses. The dominant production mechanism for such particles may
be scalar decays, as suggested by Figure \ref{fig:modeltwo}. Even if the
dark matter particles are thermal relics, their abundance now will be
affected by the nonstandard expansion rates at 
earlier times \cite{Kamionkowski:1990ni}-\cite{Pallis:2005bb}.

\begin{table}[h]
\centering
\begin{tabular}{|c|c|c|c|c|c|c|}
\hline\hline
log$_{10}$ $a $ & -42 & -40 & -38 & -36 & -34 & -32 \\ [0.5 ex]
\hline
$w(a)$ & -0.32 & -0.83 & -1 & -0.75 & -0.19 & 0.45 \\ \hline
$T$ & 0 & 2 10$^{17}$ & 3 10$^{16}$ & 10$^{17}$ & 2 10$^{16}$ & 8 10$^{14}$ \\ \hline
$\Omega_\Lambda$ & 1 & 0.997 & 0.99998 & 0.996 &
0.976 & 0.79 \\ \hline
$\Omega_r$ & 0 & 0.003 & 2 10$^{-5}$ & 0.004 & 0.024 & 0.21 \\ \hline
$\Omega_m$ & 0 & 5 10$^{-14}$ & 3 10$^{-16}$ & 5 10$^{-14}$
& 4 10$^{-13}$ & 2 10$^{-11}$   \\\hline
log$_{10}$ $a $ & -30 & -28 & -26 & -24 & -22 & -20 \\\hline
$w(a)$ & 0.90 & 0.98 & 0.64 & 0.04 & -0.58 & -0.96 \\ \hline
$T$ & 9 10$^{12}$ & 9 10$^{10}$ & 9 10$^8$ & 9 10$^{6}$ & 9 10$^{4}$ & 900  \\ \hline
$\Omega_\Lambda$ & 0.01 & 8 10$^{-7}$ & 5 10$^{-10}$ & 
2 10$^{-10}$ & 5 10$^{-7}$ & 0.07 \\\hline
$\Omega_r$ & 0.99 & 0.999998 & 0.99993 & 0.993 & 0.59 & 0.01 \\\hline
$\Omega_m$ & 7 10$^{-9}$ & 7 10$^{-7}$ & 7 10$^{-5}$
& 0.007 & 0.41 & 0.92   \\\hline
log$_{10}$ $a $ & -18 & -16 & -14 & -12 & -10 & -8 \\\hline
$w(a)$ & -0.94 & -0.52 & 0.11 & 0.69 & 0.99 & 0.87 \\ \hline 
$T$ & 220 & 170 & 17 & 0.3 & 3 10$^{-3}$ & 3 10$^{-5}$  \\ \hline
$\Omega_\Lambda$ & 0.9992 & 0.99 & 0.94 & 0.33 & 0.0002
& 2 10$^{-8}$ \\\hline
$\Omega_r$ & 0.0008 & 0.01 & 0.06 & 0.67 & 0.9998 & 0.9999  \\\hline
$\Omega_m$ & 2 10$^{-5}$ & 5 10$^{-10}$ & 3 10$^{-11}$
& 5 10$^{-9}$ & 8 10$^{-7}$ &  10$^{-4}$ \\\hline
log$_{10}$ $a $ & -6 & -4 & -2 & 0 &  &  \\\hline
$w(a)$ & 0.39 & -0.25 & -0.79 & -1.0 &  &  \\ \hline
$T$ & 3 10$^{-7}$ & 3 10$^{-9}$ & 3 10$^{-11}$ & 3 10$^{-13}$ &  &   \\ \hline
$\Omega_\Lambda$ & 10$^{-10}$ & 2 10$^{-9}$ & 
7 10$^{-6}$ & 0.71 &  &   \\\hline
$\Omega_r$ & 0.992 & 0.56 & 0.01 & 0.00005 &  &  \\\hline
$\Omega_m$ & 0.008 & 0.44 & 0.99 & 0.29 &  &  \\[1ex]
\hline\hline
\end{tabular}
\caption{Model 2: the relative density fractions of dark energy, radiation,
and matter, as a function of the scale factor. Also shown are the
temperature $T$ in GeV, and the equation of state parameter $w(a)$.}
\end{table}

\subsection{A model with many inflationary epochs}

This model is defined by:
\be
b=0.43 \; ,\quad k=0.33\;\theta (10^{-10}-a)\; ,\quad f_m = 3\times 10^{-14}\; ,
\quad a_i = 10^{-35} \; ,
\ee
where again we have used a step function $\theta (10^{-10}-a)$ to crudely
turn off the coupling of the scalar to radiation and matter at late times.

This model has many inflationary epochs, combined with strong tracking.
The horizon problem is solved because   
the ratio of the comoving
horizon to the comoving Hubble radius, as measured today,
is about 3. But this is the cumulative effect of five different
inflationary epochs.

Model 3 has inflationary epochs somewhat before and somewhat after the electroweak
phase transition. There is also a period of accelerated expansion which
occurs after BBN and ends around the time of recombination. Certainly
these effects could be important for predicting the abundances 
of WIMPs, gravitinos, modulinos and other exotic relics.

From Table 3, one notes that that radiation fraction of Model 3 at the
time of BBN is only about 0.92. This saturates the lower bound
\cite{Bean:2001wt} required for successful BBN.

\begin{figure}
\centerline{\epsfxsize 5.8 truein \epsfbox {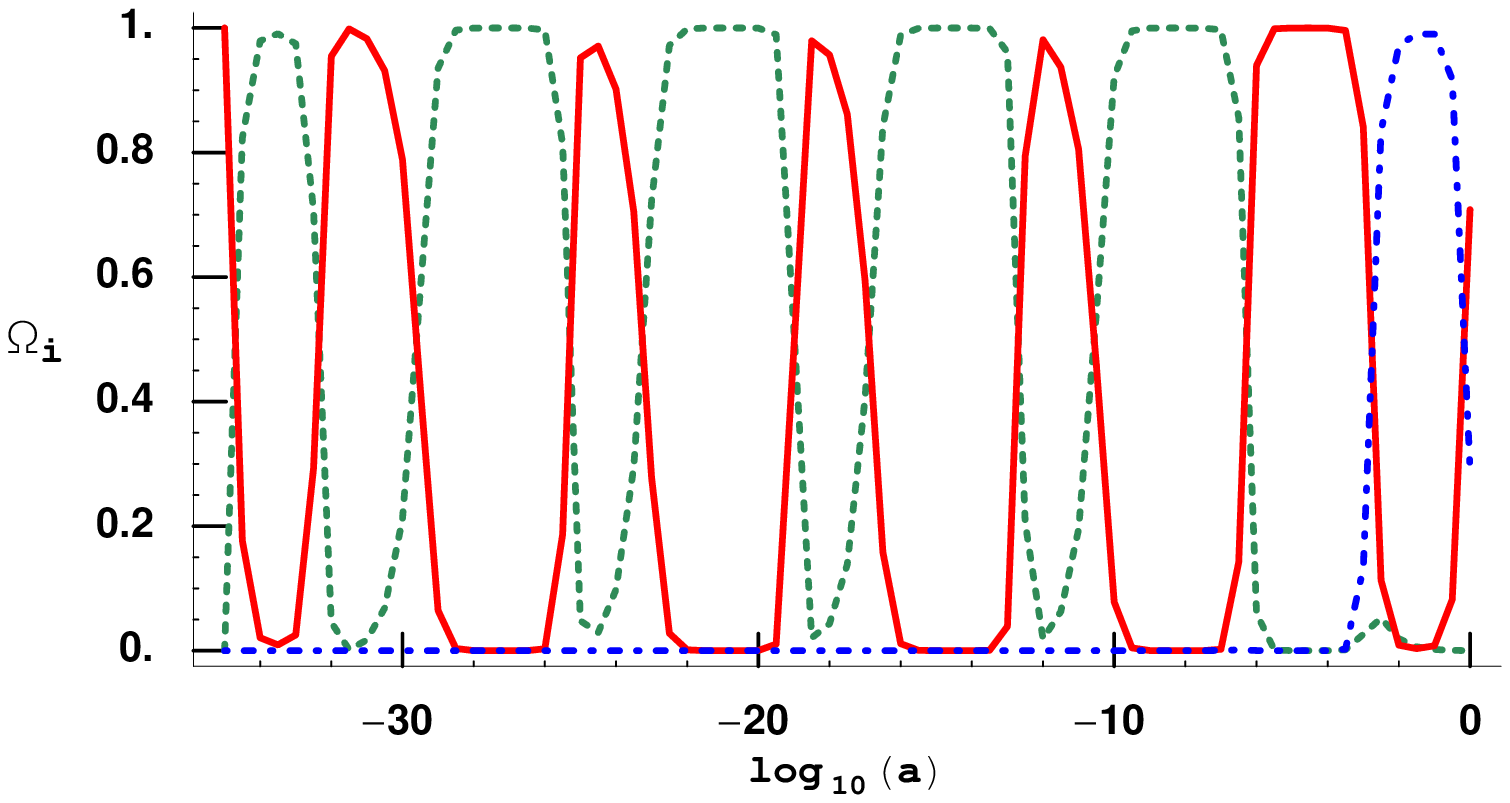}}
\caption{\label{fig:modelthree} The cosmological history of
Model 3. Shown are the relative energy density fractions
in radiation (green/dashed), matter (blue/dot--dashed), and
the noncanonical scalar (red/solid), as a function of the
logarithm of the scale factor.
\hbox to170pt{}}
\end{figure}

\begin{table}[h]
\centering
\begin{tabular}{|c|c|c|c|c|c|c|}
\hline\hline
log$_{10}$ $a $ & -34 & -32 & -30 & -28 & -26 & -24 \\ [0.5 ex]
\hline
$w(a)$ & 0.63 & -0.96 & 0.14 & 0.85 & -0.82 & -0.20 \\ \hline
$T$ & 4 10$^{16}$ & 4 10$^{14}$ & 2 10$^{14}$ & 2 10$^{12}$ & 2 10$^{10}$ & 10$^{9}$ \\ \hline
$\Omega_\Lambda$ & 0.02 & 0.95 & 0.79 & 0.0002 &
0.003 & 0.9 \\ \hline
$\Omega_r$ & 0.98 & 0.05 & 0.21 & 0.9998 & 0.997 & 0.1 \\ \hline
$\Omega_m$ & 2 10$^{-13}$ & 8 10$^{-13}$ & 10$^{-14}$ & 3 10$^{-12}$
& 3 10$^{-10}$ & 2 10$^{-12}$   \\\hline
log$_{10}$ $a $ & -22 & -20 & -18 & -16 & -14 & -12 \\\hline
$w(a)$ & 0.98 & -0.58 & -0.52 & 0.99 & -0.27 & -0.77 \\ \hline
$T$ & 2 10$^{7}$ & 2 10$^{5}$ & 8000 & 300 & 3 & 0.04 \\ \hline
$\Omega_\Lambda$ & 0.001 & 0.0003 & 0.96 & 
0.01 & 0.00005 & 0.98 \\\hline
$\Omega_r$ & 0.999 & 0.9997 & 0.04 & 0.99 & 0.99995 & 0.02 \\\hline
$\Omega_m$ & 2 10$^{-10}$ & 2 10$^{-8}$ & 6 10$^{-10}$
& 9 10$^{-9}$ & 9 10$^{-7}$ & 3 10$^{-7}$   \\\hline
log$_{10}$ $a $ & -10 & -8 & -6 & -4 & -2 & 0 \\\hline
$w(a)$ & 0.89 & 0.07 & -0.94 & 0.68 & 0.40 & -1.0 \\ \hline 
$T$ & 3 10$^{-3}$ & 3 10$^{-5}$ & 3 10$^{-7}$ & 3 10$^{-9}$ & 3 10$^{-11}$ & 3 10$^{-13}$  \\ \hline
$\Omega_\Lambda$ & 0.08 & 0.00002 & 0.94 & 0.9998 & 0.01
& 0.71 \\\hline
$\Omega_r$ & 0.92 & 0.99993 & 0.06 & 0.0001 & 0.02 & 0.00006  \\\hline
$\Omega_m$ & 5 10$^{-7}$ & 0.00005 & 0.0003
& 0.0001 & 0.97 &  0.29 \\[1ex]
\hline\hline
\end{tabular}
\caption{Model 3: the relative density fractions of dark energy, radiation,
and matter, as a function of the scale factor. Also shown are the
temperature $T$ in GeV, and the equation of state parameter $w(a)$.}
\end{table}

\section{Outlook}

We have described a simple framework for a large class of models
in which a single noncanonical scalar drives quintessential inflation.
Such models are minimal in the sense that they carry no additional
theoretical baggage beyond the standard assumptions that we have
reviewed. We have presented three examples of new models in this
class. To generate the examples, we made convenient choices of
$F(\theta )$, $V(\theta )$, and $P(X)$, and we
took the couplings of the scalar
to radiation and matter to be perturbative, which simplifies
the analysis. This framework could be made more robust by, {\it e.g.},
including the possibility of preheating, or looking at examples with
nontrivial $P(X)$. By modifying the
arbitrary forms used in (\ref{eqn:ourkin}) and (\ref{eqn:ourpot}),
it should be possible to connect our framework to a number of existing
top--down models. 

Our three examples look very different from standard inflation-assisted
$\Lambda$CDM cosmology. We would expect therefore that observational
data can discriminate among them. It may be that a more detailed
comparison with existing data is sufficient to rule out all three models.
However we would then expect that minor modifications of these models
would make at least some of them viable again.

All of this emphasizes the importance of understanding TeV
cosmology from independent physics inputs, as has been done so
successfully with MeV cosmology and Big Bang Nucleosynthesis.
A common feature of our nonstandard cosmologies is that they
affect both the electroweak phase transition (at $T\sim 100$ GeV)
and the abundance of weakly interacting dark matter components.
Most theory papers on the EWPT or dark matter abundances simply
assume standard cosmology, which is certainly dangerous. A better
strategy is to regard TeV cosmology as one of the important
{\it outputs} of particle physics. This will require
digesting the results of the  next generation of experiments
at colliders, direct dark matter searches and other experiments
and observations. 

The goal should be a ``TeV signpost'' as constraining to
cosmological model building as BBN is now. This may be the only
robust way to rule out (or rule in) the type of nonstandard
scenarios that we have presented here.

\subsection*{Acknowledgments}
The authors are grateful to Scott Dodelson, Will Kinney and
Olga Mena for useful discussions. 
JL acknowledges the support of the Aspen Center for Physics, 
as well as the hospitality of Jos\'e Bernab\'eu and the 
Universitat de Val\`encia, where this work was completed.
GB is grateful for support from the Spanish MEC and FEDER under 
Contract FPA 2005/1678,
and to the Generalitat Valenciana under Contract GV05/267.
Fermilab is operated by Universities Research Association 
Inc. under Contract No. DE-AC02-76CHO3000 with the U.S. Department
of Energy.

\newpage

% ------------------------------------------------------------------------


\begin{thebibliography}{99}

\bibitem{Spergel:2006hy}
  D.~N.~Spergel {\it et al.},
  %``Wilkinson Microwave Anisotropy Probe (WMAP) three year results:
  %Implications for cosmology,''
  arXiv:astro-ph/0603449.
  %%CITATION = ASTRO-PH 0603449;%%

\bibitem{Barenboim:2005np}
  G.~Barenboim and J.~D.~Lykken,
  %``Slinky inflation,''
  Phys.\ Lett.\ B {\bf 633}, 453 (2006)
  [arXiv:astro-ph/0504090].
  %%CITATION = ASTRO-PH 0504090;%%

%%% k essence papers

\bibitem{Chiba:1999ka}
  T.~Chiba, T.~Okabe and M.~Yamaguchi,
  %``Kinetically driven quintessence,''
  Phys.\ Rev.\ D {\bf 62}, 023511 (2000)
  [arXiv:astro-ph/9912463].
  %%CITATION = ASTRO-PH 9912463;%%

\bibitem{Armendariz-Picon:2000dh}
  C.~Armendariz-Picon, V.~F.~Mukhanov and P.~J.~Steinhardt,
  %``A dynamical solution to the problem of a small cosmological constant  and
  %late-time cosmic acceleration,''
  Phys.\ Rev.\ Lett.\  {\bf 85}, 4438 (2000)
  [arXiv:astro-ph/0004134].
  %%CITATION = ASTRO-PH 0004134;%%

\bibitem{Malquarti:2003nn}
  M.~Malquarti, E.~J.~Copeland, A.~R.~Liddle and M.~Trodden,
  %``A new view of k-essence,''
  Phys.\ Rev.\ D {\bf 67}, 123503 (2003)
  [arXiv:astro-ph/0302279].
  %%CITATION = ASTRO-PH 0302279;%%


%%% preheating papers

\bibitem{Kofman:1994rk}
  L.~Kofman, A.~D.~Linde and A.~A.~Starobinsky,
  %``Reheating after inflation,''
  Phys.\ Rev.\ Lett.\  {\bf 73}, 3195 (1994)
  [arXiv:hep-th/9405187].
  %%CITATION = HEP-TH 9405187;%%

\bibitem{Shtanov:1994ce}
  Y.~Shtanov, J.~H.~Traschen and R.~H.~Brandenberger,
  %``Universe reheating after inflation,''
  Phys.\ Rev.\ D {\bf 51}, 5438 (1995)
  [arXiv:hep-ph/9407247].
  %%CITATION = HEP-PH 9407247;%%

\bibitem{Kofman:1997yn}
  L.~Kofman, A.~D.~Linde and A.~A.~Starobinsky,
  %``Towards the theory of reheating after inflation,''
  Phys.\ Rev.\ D {\bf 56}, 3258 (1997)
  [arXiv:hep-ph/9704452].
  %%CITATION = HEP-PH 9704452;%%

\bibitem{Greene:1997fu}
  P.~B.~Greene, L.~Kofman, A.~D.~Linde and A.~A.~Starobinsky,
  %``Structure of resonance in preheating after inflation,''
  Phys.\ Rev.\ D {\bf 56}, 6175 (1997)
  [arXiv:hep-ph/9705347].
  %%CITATION = HEP-PH 9705347;%%

\bibitem{Dufaux:2006ee}
  J.~F.~Dufaux, G.~Felder, L.~Kofman, M.~Peloso and D.~Podolsky,
  %``Preheating with trilinear interactions: Tachyonic resonance,''
  arXiv:hep-ph/0602144.
  %%CITATION = HEP-PH 0602144;%%

%%%%%%%%%%%%%%%%%%

\bibitem{Tsamis:2003px}
  N.~C.~Tsamis and R.~P.~Woodard,
  %``Improved estimates of cosmological perturbations,''
  Phys.\ Rev.\ D {\bf 69}, 084005 (2004)
  [arXiv:astro-ph/0307463].
  %%CITATION = ASTRO-PH 0307463;%%

\bibitem{Grishchuk:1994sj}
  L.~P.~Grishchuk,
  %``Density perturbations of quantum mechanical origin and anisotropy of the
  %microwave background,''
  Phys.\ Rev.\ D {\bf 50}, 7154 (1994)
  [arXiv:gr-qc/9405059].
  %%CITATION = GR-QC 9405059;%%

\bibitem{Kinney:2005vj}
  W.~H.~Kinney,
  %``Horizon crossing and inflation with large eta,''
  Phys.\ Rev.\ D {\bf 72}, 023515 (2005)
  [arXiv:gr-qc/0503017].
  %%CITATION = GR-QC 0503017;%%

\bibitem{baren-kinney}
G.~Barenboim and W.~H.~Kinney, to appear shortly.


\bibitem{Linde:1993cn}
  A.~D.~Linde,
  %``Hybrid inflation,''
  Phys.\ Rev.\ D {\bf 49}, 748 (1994)
  [arXiv:astro-ph/9307002].
  %%CITATION = ASTRO-PH 9307002;%%

\bibitem{Garcia-Bellido:1996ke}
  J.~Garcia-Bellido and D.~Wands,
  %``The spectrum of curvature perturbations from hybrid inflation,''
  Phys.\ Rev.\ D {\bf 54}, 7181 (1996)
  [arXiv:astro-ph/9606047].
  %%CITATION = ASTRO-PH 9606047;%%

\bibitem{Kinney:1997ne}
  W.~H.~Kinney,
  %``A Hamilton-Jacobi approach to non-slow-roll inflation,''
  Phys.\ Rev.\ D {\bf 56}, 2002 (1997)
  [arXiv:hep-ph/9702427].
  %%CITATION = HEP-PH 9702427;%%

\bibitem{Garcia-Bellido:1997wm}
  J.~Garcia-Bellido and A.~D.~Linde,
  %``Preheating in hybrid inflation,''
  Phys.\ Rev.\ D {\bf 57}, 6075 (1998)
  [arXiv:hep-ph/9711360].
  %%CITATION = HEP-PH 9711360;%%


% recontructing the inflaton potential:

\bibitem{Copeland:1993zn}
  E.~J.~Copeland, E.~W.~Kolb, A.~R.~Liddle and J.~E.~Lidsey,
  %``Reconstructing the inflaton potential: Perturbative reconstruction to
  %second order,''
  Phys.\ Rev.\ D {\bf 49}, 1840 (1994)
  [arXiv:astro-ph/9308044].
  %%CITATION = ASTRO-PH 9308044;%%

\bibitem{Liddle:1994cr}
  A.~R.~Liddle and M.~S.~Turner,
  %``Second order reconstruction of the inflationary potential,''
  Phys.\ Rev.\ D {\bf 50}, 758 (1994)
  [Erratum-ibid.\ D {\bf 54}, 2980 (1996)]
  [arXiv:astro-ph/9402021].
  %%CITATION = ASTRO-PH 9402021;%%

\bibitem{Copeland:1998fz}
  E.~J.~Copeland, I.~J.~Grivell, E.~W.~Kolb and A.~R.~Liddle,
  %``On the reliability of inflaton potential reconstruction,''
  Phys.\ Rev.\ D {\bf 58}, 043002 (1998)
  [arXiv:astro-ph/9802209].
  %%CITATION = ASTRO-PH 9802209;%%

\bibitem{Grivell:1999wc}
  I.~J.~Grivell and A.~R.~Liddle,
  %``Inflaton potential reconstruction without slow-roll,''
  Phys.\ Rev.\ D {\bf 61}, 081301 (2000)
  [arXiv:astro-ph/9906327].
  %%CITATION = ASTRO-PH 9906327;%%

\bibitem{Easther:2002rw}
  R.~Easther and W.~H.~Kinney,
  %``Monte Carlo reconstruction of the inflationary potential,''
  Phys.\ Rev.\ D {\bf 67}, 043511 (2003)
  [arXiv:astro-ph/0210345].
  %%CITATION = ASTRO-PH 0210345;%%

\bibitem{Kinney:2003uw}
  W.~H.~Kinney, E.~W.~Kolb, A.~Melchiorri and A.~Riotto,
  %``WMAPping inflationary physics,''
  Phys.\ Rev.\ D {\bf 69}, 103516 (2004)
  [arXiv:hep-ph/0305130].
  %%CITATION = HEP-PH 0305130;%%

\bibitem{Kadota:2005hv}
  K.~Kadota, S.~Dodelson, W.~Hu and E.~D.~Stewart,
  %``Precision of inflaton potential reconstruction from CMB using the  general
  %slow-roll approximation,''
  Phys.\ Rev.\ D {\bf 72}, 023510 (2005)
  [arXiv:astro-ph/0505158].
  %%CITATION = ASTRO-PH 0505158;%%

\bibitem{Cline:2006db}
  J.~M.~Cline and L.~Hoi,
  %``Inflationary Potential Reconstruction for a WMAP Running Power Spectrum,''
  arXiv:astro-ph/0603403.
  %%CITATION = ASTRO-PH 0603403;%%

%%%%%%%%%%%%%%%%%%%%%%


\bibitem{Barenboim:2004kz}
  G.~Barenboim, O.~Mena and C.~Quigg,
  %``Undulant universe,''
  Phys.\ Rev.\ D {\bf 71}, 063533 (2005)
  [arXiv:astro-ph/0412010].
  %%CITATION = ASTRO-PH 0412010;%%

\bibitem{Barenboim:2005wx}
  G.~Barenboim, O.~Mena Requejo and C.~Quigg,
  %``Observational constraints on undulant cosmologies,''
  JCAP {\bf 0604}, 008 (2006)
  [arXiv:astro-ph/0510178].
  %%CITATION = ASTRO-PH 0510178;%%

\bibitem{Carroll:1998zi}
  S.~M.~Carroll,
  %``Quintessence and the rest of the world,''
  Phys.\ Rev.\ Lett.\  {\bf 81}, 3067 (1998)
  [arXiv:astro-ph/9806099].
  %%CITATION = ASTRO-PH 9806099;%%

\bibitem{Feldman:2006wg}
  B.~Feldman and A.~Nelson,
  %``New regions for a chameleon to hide,''
  arXiv:hep-ph/0603057.
  %%CITATION = HEP-PH 0603057;%%

%%%%  tracking

\bibitem{Zlatev:1998tr}
  I.~Zlatev, L.~M.~Wang and P.~J.~Steinhardt,
  %``Quintessence, Cosmic Coincidence, and the Cosmological Constant,''
  Phys.\ Rev.\ Lett.\  {\bf 82}, 896 (1999)
  [arXiv:astro-ph/9807002].
  %%CITATION = ASTRO-PH 9807002;%%

%%%% spectral indices

\bibitem{Nunes:2002wz}
  N.~J.~Nunes and E.~J.~Copeland,
  %``Tracking quintessential inflation from brane worlds,''
  Phys.\ Rev.\ D {\bf 66}, 043524 (2002)
  [arXiv:astro-ph/0204115].
  %%CITATION = ASTRO-PH 0204115;%%

%%% EW baryogenesis papers

\bibitem{Joyce:1997fc}
  M.~Joyce and T.~Prokopec,
  %``Turning around the sphaleron bound: Electroweak baryogenesis in an
  %alternative post-inflationary cosmology,''
  Phys.\ Rev.\ D {\bf 57}, 6022 (1998)
  [arXiv:hep-ph/9709320].
  %%CITATION = HEP-PH 9709320;%%

\bibitem{Trodden:1998ym}
  M.~Trodden,
  %``Electroweak baryogenesis,''
  Rev.\ Mod.\ Phys.\  {\bf 71}, 1463 (1999)
  [arXiv:hep-ph/9803479].
  %%CITATION = HEP-PH 9803479;%%

%%%%%%%%%%%%
%%% change relic density by nonstandard cosmology

\bibitem{Kamionkowski:1990ni}
  M.~Kamionkowski and M.~S.~Turner,
  %``Thermal Relics: Do We Know Their Abundances?,''
  Phys.\ Rev.\ D {\bf 42}, 3310 (1990).
  %%CITATION = PHRVA,D42,3310;%%

\bibitem{Profumo:2003hq}
  S.~Profumo and P.~Ullio,
  %``SUSY dark matter and quintessence,''
  JCAP {\bf 0311}, 006 (2003)
  [arXiv:hep-ph/0309220].
  %%CITATION = HEP-PH 0309220;%%

\bibitem{Catena:2004ba}
  R.~Catena, N.~Fornengo, A.~Masiero, M.~Pietroni and F.~Rosati,
  %``Dark matter relic abundance and scalar-tensor dark energy,''
  Phys.\ Rev.\ D {\bf 70}, 063519 (2004)
  [arXiv:astro-ph/0403614].
  %%CITATION = ASTRO-PH 0403614;%%

\bibitem{Pallis:2005bb}
  C.~Pallis,
  %``Kination dominated reheating and cold dark matter abundance,''
  arXiv:hep-ph/0510234.
  %%CITATION = HEP-PH 0510234;%%

%%%%%%%%%%%%%%

\bibitem{Bean:2001wt}
  R.~Bean, S.~H.~Hansen and A.~Melchiorri,
  %``Early universe constraints on dark energy,''
  Phys.\ Rev.\ D {\bf 64}, 103508 (2001)
  [arXiv:astro-ph/0104162].
  %%CITATION = ASTRO-PH 0104162;%%

\bibitem{Alabidi:2006qa}
  L.~Alabidi and D.~H.~Lyth,
  %``Inflation models after WMAP year three,''
  arXiv:astro-ph/0603539.
  %%CITATION = ASTRO-PH 0603539;%%

\end{thebibliography}
\end{document}